\documentclass[aps,prl,twocolumn,showpacs,groupedaddress,amsmath,amssymb,floatfix,tightenlines]{revtex4-1}
\usepackage{graphicx}
\usepackage[T1]{fontenc}

\usepackage{graphics}

\usepackage{color}
\usepackage{multirow}

\begin{document}


\title{The Implications of Grain Size Variation in Magnetic Field Alignment of Block Copolymer Blends}


\author{Yekaterina Rokhlenko}

\affiliation{Department of Chemical Engineering, Yale University, New Haven CT 06511}

\author{Pawe{\l} W. Majewski}
\affiliation{Center for Functional Nanomaterials, Brookhaven National Lab, Upton NY 11973}
\affiliation{Department of Chemistry, University of Warsaw, Warsaw, Poland, 02093}

\author{Steven R. Larson}
\author{Padma Gopalan}
\affiliation{Department of Materials Science and Engineering, University of Wisconsin, Madison WI 53706}

\author{Kevin G. Yager}
\affiliation{Center for Functional Nanomaterials, Brookhaven National Lab, Upton NY 11973}

\author{Chinedum O. Osuji}
\email{chinedum.osuji@yale.edu}
\affiliation{Department of Chemical Engineering, Yale University, New Haven CT 06511}

\date{\today}

\begin{abstract}
Recent experiments have highlighted the intrinsic magnetic anisotropy in coil-coil diblock copolymers, specifically in poly(styrene-\textit{block}-4-vinylpyridine) (PS-\textit{b}-P4VP), that enables magnetic field alignment at field strengths of a few tesla. We consider here the alignment response of two low molecular weight (MW) lamallae-forming PS-\textit{b}-P4VP systems. Cooling across the disorder-order transition temperature ($\mathrm{T_{odt}}$) results in strong alignment for the higher MW sample (5.5K), whereas little alignment is discernible for the lower MW system (3.6K). This disparity under otherwise identical conditions of field strength and cooling rate suggests that different average grain sizes are produced during slow cooling of these materials, with larger grains formed in the higher MW material. Blending the block copolymers results in homogeneous samples which display $\mathrm{T_{odt}}$, d-spacings and grain sizes that are intermediate between the two neat diblocks. Similarly, the alignment quality displays a smooth variation with the concentration of the higher MW diblock in the blends and the size of grains likewise interpolates between limits set by the neat diblocks, with a factor of 3.5X difference in the grain size observed in high vs low MW neat diblocks. These results highlight the importance of grain growth kinetics in dictating the field response in block copolymers and suggests an unconventional route for the manipulation of such kinetics.
\end{abstract}

%

\maketitle


\section{Introduction}

Magnetic fields can be used to direct the self-assembly of block copolymers (BCPs) in a facile manner under appropriate conditions \cite{Osuji2004,Tao2007, Gopinadhan2010,Gopinadhan2012}. Thermodynamically, it is the combination of grain size, field strength, and magnetic susceptibility anisotropy that leads to magnetic field alignment in a variety of BCP systems. The driving force for alignment of an anisotropic object is its angle dependent magnetostatic energy density $\epsilon_m$ defined in Eq. \ref{eq:Em},
\begin{eqnarray}
\label{eq:Em}
\epsilon_m&=&\frac{-B^2}{2\mu_0}\left(\chi_{\parallel}\cos^2\varphi+\chi_{\perp}\sin^2\varphi\right) \\
\label{eq:Em2}
\Delta\epsilon_m&=&\frac{-\Delta\chi B^2}{2\mu_0}
\end{eqnarray}

where $\chi_{\parallel}$ and $\chi_{\perp}$ are the magnetic susceptibilities parallel and perpendicular to the axis of highest rotational symmetry, $B$ is the field strength, and $\mu_0$ is the permeability of free space. The magnetostatic energy density difference between orthogonal alignments is given in Eq. \ref{eq:Em2}, where $\Delta\chi=\chi_{\parallel}-\chi_{\perp}$. Alignment of an object with volume $V_g$ is possible when the extensive quantity, the magnetostatic energy difference between orthogonal alignments, $\Delta E_m$, exceeds thermal energy $k_{B}T$, Eq. \ref{eq:EmVg}. For BCPs, $V_g=\xi^3$ is the volume of a grain with characteristic dimension $\xi$.

\begin{eqnarray}
\label{eq:EmVg}
|\Delta E_m|=|\Delta\epsilon_m| V_g\gg k_BT
\end{eqnarray}

Mesogen attachment to polymer backbones to yield liquid crystalline (LC) BCPs has been a typical means of providing a sufficiently large $\Delta\chi$ ($\approx 10^{-6}$ in dimensionless SI volume units) for field alignment at reasonable field strengths\cite{Ferri1998,Hamley2004,Gopinadhan2012,Majewski2012,Tran2013}, for grain sizes of hundreds of nm. Recently, it has been shown that a simple coil-coil BCP (i.e. non-LC BCP), PS-\textit{b}-P4VP, displays sufficiently large grain sizes and intrinsic magnetic anisotropy to be well-aligned by a magnetic field at appropriate molecular weights.\cite{Rokhlenko2015} The intrinsic magnetic anisotropy originates in the correlation of the orientations of end-end vectors necessitated by the localization of block junctions at the microdomain interface. The orientation distribution of the end-end vectors has its maximum along the intermaterial dividing surface (IMDS) normal, which, for PS-\textit{b}-P4VP, imparts a net magnetic anisotropy of $\Delta\chi\approx$ 1.6 x $10^{-8}$ for the system overall. Based on this analysis, compared to LC BCPs, coil-coil BCPs have $\Delta\chi$ that is roughly 2 orders of magnitude lower. Thus the expected minimum grain size for alignment is about 5 times larger ($10^{2/3}$), as suggested by the linear relationship between the magnetostatic energy, $\Delta E_m$, and volume, $V_g=\xi^3$ in Eq. \ref{eq:EmVg}.

As a BCP is cooled through $\mathrm{T_{odt}}$, the blocks begin to microphase separate and the system becomes increasingly magnetically anisotropic. This is accompanied by a sharp decrease in the mobility of the chains, or a rapid increase in viscosity. The increasing viscosity begins to kinetically prohibit alignment, while the increasing magnetic anisotropy favors alignment thermodynamically. Thus, the degree of alignment is a sensitive function of undercooling, and the cooling rate through $\mathrm{T_{odt}}$ plays an important role in grain alignment - slower cooling improves alignment until a thermodynamic limit is reached \cite{Gopinadhan2013}. Additionally, slower cooling can have an indirect effect on the energetics of field alignment through an increased grain size.

Here we examine differences in alignment of two low MW PS-\textit{b}-P4VP BCPs at the same cooling rate and field strength. Blending the BCPs produces samples with intermediate $\mathrm{T_{odt}}$, d-spacings and grain sizes. Concurrently, the alignment quality varies with composition, and is related to grain size - samples with larger grains display better alignment. The contrasting alignment quality of the two neat diblocks, which differ in grain size by a factor of 3.5, shows that changes in grain size can have profound effects on field-induced orientational order. Blending of these diblocks provides an unconventional route to vary grain size in block copolymers.

\begin{figure}[t]
\includegraphics [width=80mm,scale=1]{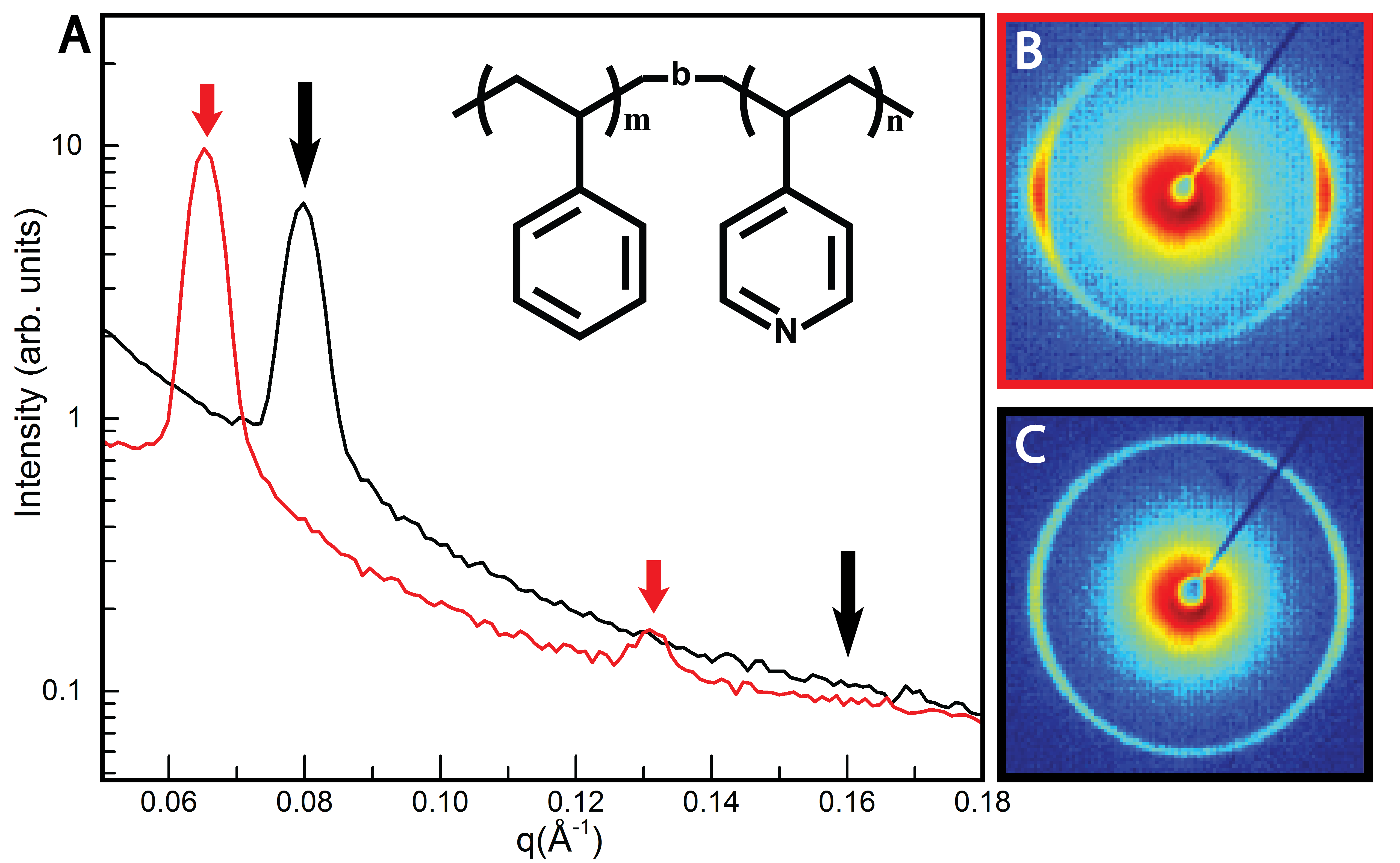}
  \caption{(a) SAXS data for 3.6K (black) and 5.5K (red) samples. Arrows indicate primary peak, and expected location of second order peak based for lamellar morphologies. Representative TEM images are available in the Supporting Information. Inset: Chemical structure of PS-b-P4VP. (b,c) Resulting 2D diffractograms after cooling 5.5K and 3.6K samples, respectively, at 0.3 K/min. in a 6 T field. The field direction is vertical in the plane of the diffractograms, i.e. along the meridional line.}
  \label{Fig1}
\end{figure}
The samples under investigation are lamellae-forming PS-b-P4VP with MWs of 3.6 kg/mol. (K) and 5.5K, with P4VP weight fractions, f$_{P4VP}=0.51$ and $0.47$, and d-spacings of 7.9 and 9.5 nm, respectively (Figure \ref{Fig1}). Slow cooling the 5.5K sample at 0.3 K/min. under a 6 T field from the disordered melt (T>$\mathrm{T_{odt}}$) results in pronounced alignment of the lamellar normals perpendicular to the applied field, as inferred by small-angle X-ray scattering (SAXS) (Figure \ref{Fig1}B). The 3.6K sample shows minimal alignment when subjected to the same protocol, as demonstrated by the isotropic ring-like scattering pattern (Figure \ref{Fig1}C).
\begin{figure}[h]
\includegraphics[width=80mm, scale=1]{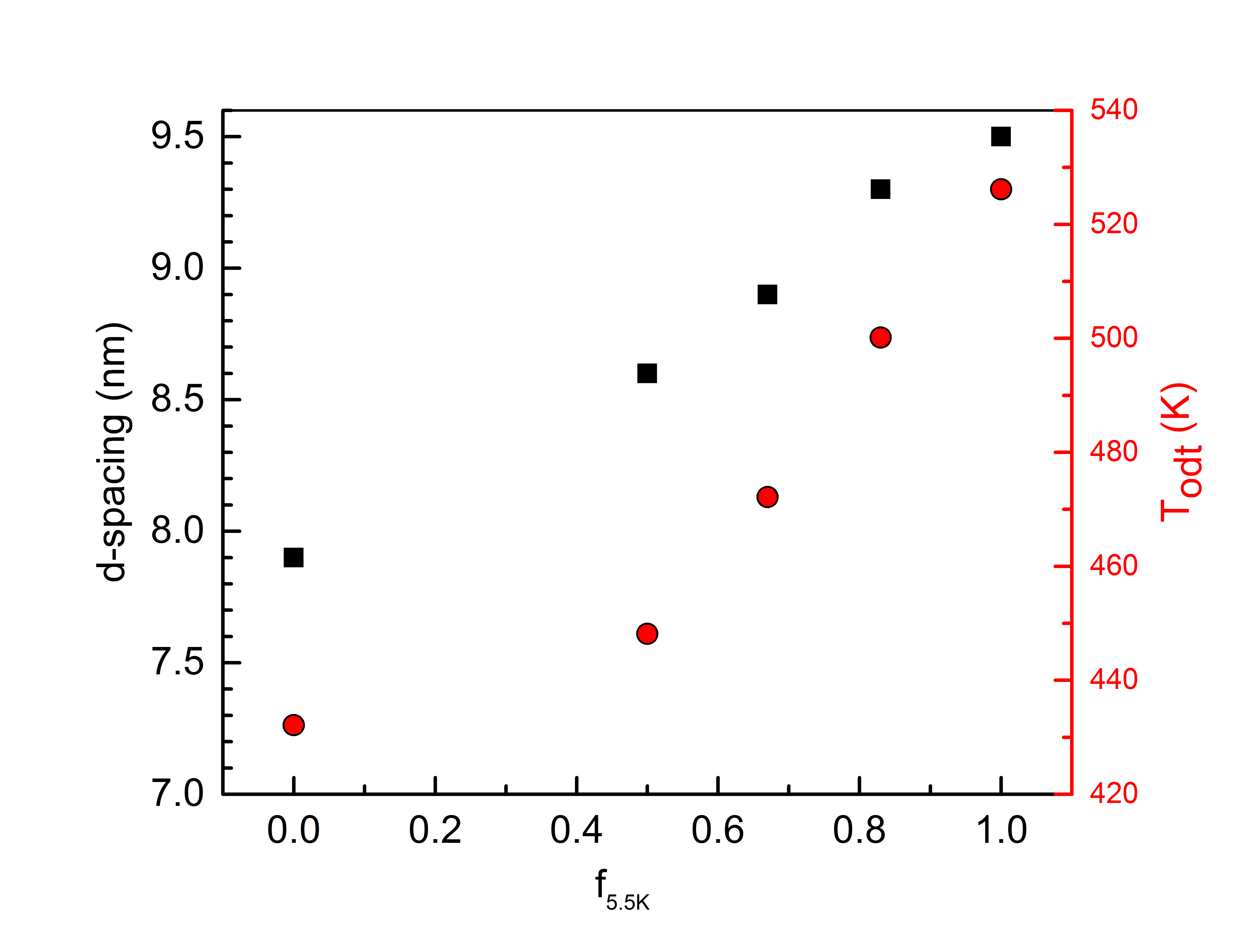}
  \caption{Experimentally determined d-spacings (black squares) and $\mathrm{T_{odt}}$ values (red circles) as a function of the weight fraction of 5.5K material.}
  \label{Fig2}
\end{figure}
Given the identical cooling rate and chemistry of the two neat BCPs, it is likely that the differences in alignment behavior are due to variations in grain size, originating from differing nucleation and growth kinetics near $\mathrm{T_{odt}}$. We considered the possibility that blends of the two materials would interpolate their neat field alignment responses, and speculated that any grain size differences would also be reflected in this alignment response.

Blending with homopolymer \cite{Matsen1995,Floudas1997,Perlich2007} or BCPs with different MWs\cite{hashimoto1993ordered,Kane1996,Zhang2011} has been used to tailor both the morphology and microdomain size of BCPs. For binary BCP blends, complete miscibility is observed for MW ratios less than $\approx$5. We observe a roughly linear variation of d-spacing with blend ratio, as expected for strongly segregating systems such as PS-\textit{b}-P4VP (Figure \ref{Fig2}). BCP ordering at finite MWs is influenced by fluctuation effects, with ordered states produced by a fluctuation-induced first order transition\cite{fredrickson1987fluctuation}, rather than via the critical point anticipated by the original mean-field theory\cite{Leibler1980}. The lamellar order-disorder transition (ODT) occurs at $(\chi_e N)\mathrm{_{odt}}=10.495+41.0\bar{N}^{-1/3}$, where $\chi_e$ is the effective Flory interaction parameter, $N$ is the (statistical) degree of polymerization, and $\bar{N}=N(cb^3)^2$ is the invariant degree of polymerization, with $b$ the statistical segment length and $c$ the monomer concentration. The effective interaction parameter nominally scales inversely proportional to temperature, $\chi_e\sim 1/T$ and so for a MW independent $\chi_e$, one expects $\mathrm{T_{odt}}\sim N$. In reality, $\chi_e = A + B/T$, where $A$ and $B$ are constants relating to entropic and enthalpic characteristics of the interaction parameter, respectively. A linear dependence on composition is expected if composition variation is interpreted as an effective molecular weight dependence, $\bar{M}_{n,{\alpha\beta}}$ and only if the entropic term $A$ is negligible.

\begin{figure*}
\includegraphics[width=120mm, scale=1]{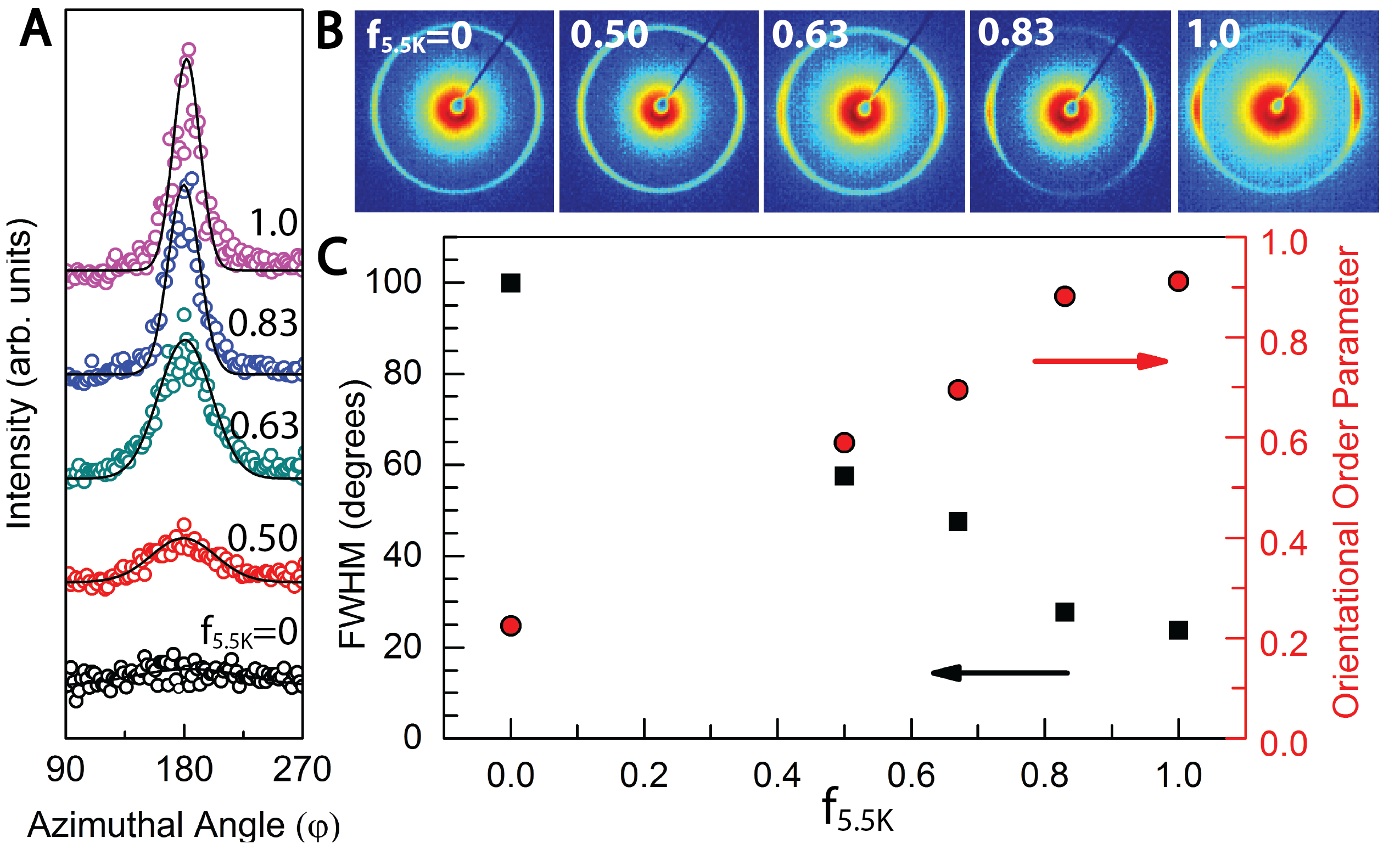}
  \caption{(a)Radially-integrated SAXS data for samples after cooling through $\mathrm{T_{odt}}$ at 0.3 K/min. under a 6 T field. Amplitude-constrained Gaussian fits are used to model the data. (b) 2D SAXS diffractrograms corresponding to 1D data shown in (a). (c) FWHM of the peaks shown in (a) extracted using the fitted Gaussian curves (black squares) and calculated orientational order parameters $\langle P_2\rangle$ based on FWHM values (red circles).}
\label{Fig3}
\end{figure*}

Figure \ref{Fig2} shows $\mathrm{T_{odt}}$ values determined from SAXS as described in the Supporting Information. Though the blend $\mathrm{T_{odt}}$ values interpolate between those of the neat diblocks ($f_{5.5K}$=0 and $f_{5.5K}$=1), there appears to be a deviation from linearity for $f_{5.5K}$=0, which suggests that some of the underlying assumptions about $\chi_e$ and the magnitude of the entropic term $A$ may be incorrect.

We studied the alignment quality of the blends and compared them to the neat diblocks using \textit{in-situ} SAXS. The resulting 2D diffractograms after cooling from $\mathrm{T_{odt}+10}$K to $\mathrm{T_{odt}-20}$K at 0.3 K/min at 6 T are shown in Figure \ref{Fig3}. The orientational order of the lamellar microdomains improves markedly as the weight fraction of 5.5K material in the sample, $f_{5.5K}$, increases, as evidenced by the increasing intensity concentration azimuthally. The azimuthal intensity dependence $I(\varphi)$ reflects the probability $p(\varphi,B)$ of observing lamellar normals at a given angle $\varphi$ with respect to the applied field direction. Because $p(\varphi,B)$ is based on Boltzmann factors incorporating the angle dependent magnetostatic energy $E_m$, the full width at half maximum (FWHM) from Gaussian fits of $I(\varphi)$ can be used to calculate the orientation distribution coefficient $\langle P_2\rangle$ of the lamellar microdomains by integration, Eq. \ref{eq:P2_energy2}. Figure \ref{Fig3} shows $I(\varphi)$ for the five samples, and the corresponding FWHM and $\langle P_2\rangle$ values. The degree of orientation as captured by $\langle P_2\rangle$ increases roughly linearly with $f_{5.5K}$.

\begin{eqnarray}
\label{eq:P2_energy1}
 p(\varphi,B)=\frac{e^{-E_m/kT}\sin\varphi\,d\varphi}{\int^{\pi}_0e^{-E_m/kT}\sin\varphi\,d\varphi} \\
 \label{eq:P2_energy2}
\langle P_2\rangle=\frac{\int^{\pi}_0 (\frac{3}{2}\cos^2\varphi-\frac{1}{2})\,e^{-E_m/kT}\,\sin\varphi\,d\varphi}{\int^{\pi}_0e^{-E_m/kT}\sin\varphi\,d\varphi}
\end{eqnarray}

\begin{figure}[h]
\includegraphics[width=80mm, scale=1]{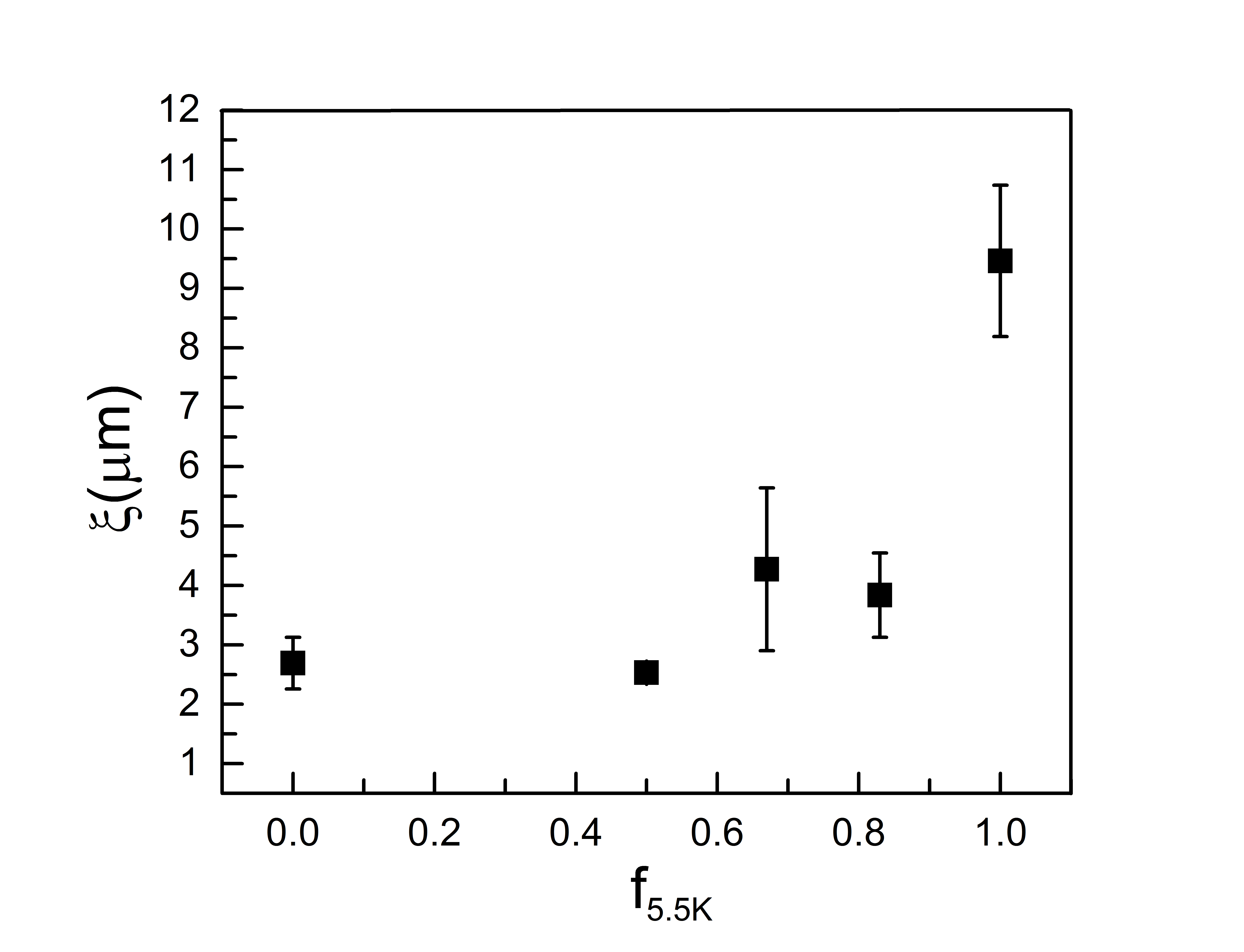}
  \caption{Calculated characteristic grain sizes based on variance scattering analysis as a function of $f_{5.5K}$. The error bars represent standard deviations from 12 independently measured spots.}
  \label{Fig4}
\end{figure}

We investigated grain size differences across our samples using a recently developed `variance scattering' technique \cite{Yager2014}, which involves computing the standard deviation of $I(\varphi)$ of the primary scattering peak, from which the number of independent scatterers, and thus the characteristic grain dimension $\xi$ can be calculated (details in Supporting Information). For this grain-size estimate, non-aligned (\textit{i.e.} isotropic) samples were prepared by cooling samples through ODT at 0.3 K/min. without any field applied. The results of our analysis (Figure \ref{Fig4}) show that there is a three-fold difference between the grain sizes of the neat BCPs at room temperature. This difference is consistent with qualitative assessments of TEM images of bulk samples (Supporting Information).

During our experiment, grains of microphase separated lamellae begin to nucleate as the samples are cooled slowly through ODT. Past studies have highlighted the connections between the cooling rate and extent of undercooling, and the resulting grain sizes in BCPs. Yager \textit{et al}. showed that there was a significant correlation between faster quenching and smaller grain size for a small molecule surfactant hexagonal mesophase \cite{Yager2014}. In isothermal experiments at fixed undercoolings, Russell \textit{et al}. observed that larger grains are produced by annealing films at smaller undercoolings, \textit{i.e.} closer to $\mathrm{T_{odt}}$, for thermally annealed cylindrical microdomains.\cite{Gu2016}. Studies of grain growth kinetics have been conducted by Balsara \textit{et al.}\cite{Dai1996,balsara1998identification,Kim2001} and Lodge \textit{et al}.\cite{chastek2004grain,chastek2006grain}. They observed growth velocities which scaled with the undercooling in good agreement with ordered front propagation velocities predicted by the theory of Goveas and Milner \cite{Goveas1997}.

Within classical nucleation theory, for homogeneous nucleation, the expected characteristic grain size $\langle \xi \rangle$ is a simple function of grain growth velocity $v$ and nucleation rate $I$, $\langle \xi \rangle \approx (v/I)^{1/4}$ \cite{avrami1940kinetics,Axe1986}. Analytical expressions for both the growth and nucleation rates following the treatment of Fredrickson and Binder\cite{Binder1989} are shown in Eq. \ref{eq:v} and Eq. \ref{eq:I}. Here, $\chi_{t}$ is the Flory interaction parameter at the coexistence between the lamellar and disordered phases, $R_g\sim N^{1/2}$ is the radius of gyration, $\delta=(\chi N-\chi_t N)/\chi_t N$ is a dimensionless undercooling, $\tau_R$ is the inverse susceptibility of the disordered phase \cite{fredrickson1987fluctuation}, $\tau_d$ is the terminal chain relaxation time, and $\Delta F^*$ is the nucleation barrier. For symmetric diblock copolymers subjected to shallow quenches, the nucleation barrier $\Delta F^*/k_BT\sim \bar{N}^{-1/3}\delta^{-2}$.

\begin{eqnarray}
\label{eq:v}
v&\sim&\frac{\chi_{t}N R_g \delta}{\tau_{R}^{1/2}\tau_{d}} \\
\label{eq:I}
I&\sim&\frac{\tau_{R}^{5/2}}{\tau_{d} R_g^{1/3}} \exp\left[-\frac{\Delta F^*}{k_BT}\right]]\\
\label{eq:eps}
\langle \xi \rangle &\sim& (\chi_tN)\delta N^{3/4}\exp\left[\frac{\Delta F^*}{4k_BT}\right]
\end{eqnarray}

On the basis of this theoretical treatment, the average grain size is expected to scale as shown in Eq. \ref{eq:eps}. This treatment assumes perfectly monodisperse chains of a symmetric BCP in which nuclei are formed solely by homogeneous nucleation at a fixed temperature. In the present case, given the non-isothermal ordering, the finite polydispersity of the BCPs, and their proximity in molecular weight, we consider it unlikely that the observed differences can be rationalized by classic nucleation and growth theories interpreted in the context of molecular weight effects as captured in Eq. \ref{eq:eps}. We speculate, instead, that the differences in grain size may originate from a molecular weight dependence of the Flory interaction parameter $\chi$, in addition to MW and $\chi$ dependence of the critical nucleation barrier, or as a result of heterogeneous nucleation.

Attempts were made to alter the grain size and alignment quality of the neat BCPs by controlling heterogeneous nucleation rates through deliberate seeding of samples with nanoparticles (expected to produce smaller grains in $f_{5.5K}=1$) and filtration with a 0.2 $\mu$m selective membrane to removal of any inadvertent particulate impurities (expected to produce larger grains in $f_{5.5K}=0$). These attempts however were unsuccessful. It is possible that heterogeneous nucleation promoters present in the samples were too small to be removed by the filter, or too numerous to have their effects swamped by the added nanoparticles. The variance scattering technique gives $\xi(f_{5.5K}=1)$ = 3.5$\xi(f_{5.5K}=0)$, and because $\langle\xi\rangle \sim n^{-1/3}$, where $n$ is the number density of heterogeneous nucleation sites, this difference implies a $\approx$45X change in the density of heterogeneous nucleation sites between the two blend components. If we postulate that the $f_{5.5K}$=1 sample has very few heterogeneous nucleation sites while the $f_{5.5K}$=0 sample has $\approx$ 45X more, and that the blends contain interpolating quantities based on their compositions, we can estimate, for example, that $n(f_{5.5K}=0.83)\approx 0.20n(f_{5.5K}=0)$ and thus we would expect $\xi(f_{5.5K}=0.83)\approx 1.7\xi(f_{5.5K}=0)$. From variance scattering we observe $\xi(f_{5.5K}=0.83) = 1.4\xi(f_{5.5K}=0)$. The similarity between these two ratios (1.7 and 1.4) suggests that heterogeneous nucleation dominated by nucleation sites contributed by the low MW sample could be the reason for the observed differences in grain sizes in the blends.

\begin{figure}[h]
\includegraphics
[width=80mm, scale=1]{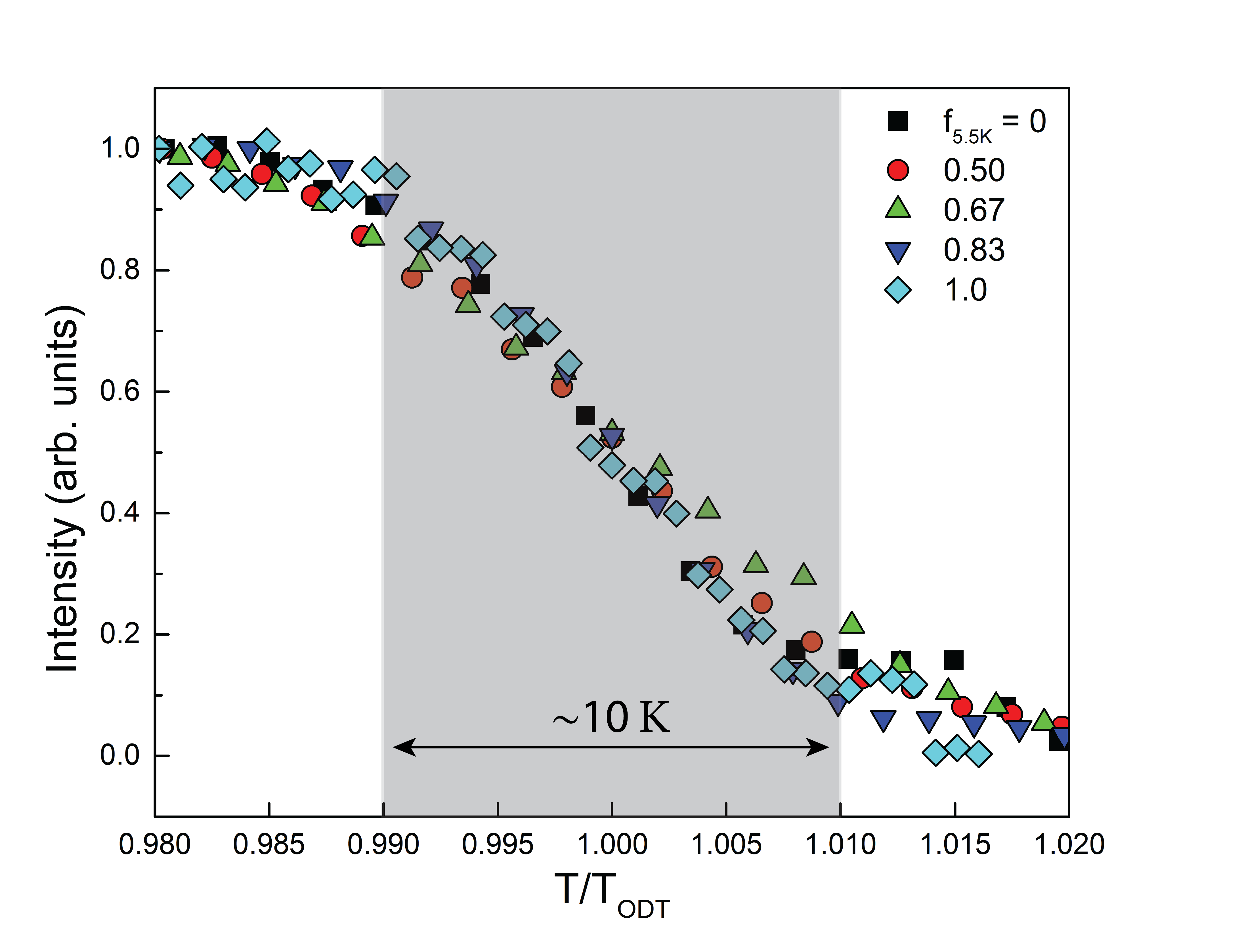}
  \caption{Reduced temperature dependence of primary Bragg peak intensity. The near-overlap of the data and corresponding near-identical width and shape of the ODT window suggest there is no significant difference in the ordering kinetics of the samples.}
  \label{Fig5}
\end{figure}

Finally, we investigated whether changes in grain size due to differing ordering kinetics could be induced by differences in the shape or width of the disorder-order transition ($\Delta\mathrm{T_{odt}}$). Specifically, we hypothesized that subtle differences in the polydispersity of the two neat diblock samples may underpin the observed differences in the grain sizes produced during ordering. Analysis of matrix-assisted laser desorption/ionization (MALDI) spectra (Supporting Information) indicates that both neat diblocks have \DJ$\approx$1.04. Additionally, plots of primary scattering peak intensity as a function of reduced temperature ($T/\mathrm{T_{odt}}$) show no discernible differences in $\Delta \mathrm{T_{odt}}$ for all five samples (Figure \ref{Fig5}), suggesting no significant differences in ordering kinetics, even for the blends which have increased polydispersity compared to the neat samples. A similar observation was reported for binary blends of PS-\textit{b}-PI where increases in polydispersity due to blending resulted in negligible effects on the width of the ordering transition, $\Delta \mathrm{T_{odt}}$\cite{Stamm1996}. We therefore conclude that the observed differences in grain size are not due to polydispersity effects.

Though we can rationalize that heterogeneous nucleation may be responsible for the grain size differences in our five samples, it is important to note that in our comparison of the  driving force for alignment for homogenous nucleation in the 3.6K and 5.5K samples, we assume that $\chi_e$ is the same in both systems, given negligible polydispersity differences and identical chemistry. Recent work, both theoretical and experimental, has shown that at low N, $\chi_e$ can be appreciably higher than for the $N \rightarrow\infty$ case, possibly due to the significance of end group effects\cite{Bates2015}. Given the low molecular weight of both samples($\approx $ 30-50 monomer units), it is likely that $\chi_e$ is a function of $N$. We expect that the ramifications of a larger $\chi_e$ may be non-trivial and that the overall effect of a larger $\chi_e$ at smaller $N$ on grain growth kinetics may be convoluted and not readily anticipated by a straightforward application of existing theory.

The field-dependent orientation distribution coefficient for the 5.5K sample, $\langle P_2 \rangle\approx$ 0.9 (red circle at $f_{5.5K}$=1 in Figure \ref{Fig3}C), yields an estimate of $\approx$ 1.2 $\mu$m for the characteristic grain size that would, at a steady state, reproduce the orientation distributions measured in aligned samples.\cite{Rokhlenko2015} Mirroring the 3.5-fold difference in grain size between the two neat diblocks calculated in this work by the variance scattering method, we can assume that the 3.6K sample has grains of $1.2/3.5 \approx$ 0.35 $\mu$m at steady state, which, from previous work \cite{Rokhlenko2015}, is predicted to give an orientational order parameter of $\langle P_2\rangle\approx$ 0.2 at 6 T.  This is comparable to what is observed experimentally for the 3.6K sample (red circle at $f_{5.5K}$=0 in Figure \ref{Fig3}C), which suggests that grain size differences alone can account for the discrepancy in alignment in these two neat diblock systems. While it may appear that there is little change in grain size between $f_{5.5K}=0$ and $f_{5.5K}=0.83$ in Figure \ref{Fig4}, the grain size in fact changes by a factor of 1.5X. Although modest, this change is expected to manifest itself in the observed alignment quality due to the strong cubic dependence of magnetostatic energy density on $\xi$ (Eq.\ref{eq:EmVg}).

In conclusion, this study relates an observed difference in alignment of two PS-\textit{b}-P4VP BCPs with a three-fold difference in grain size. The disparity in grain size contributes to markedly different alignment behaviors. The orientational order parameters are 0.2 and 0.9 for the lower and higher MW materials, respectively. There are limited ways to systematically independently manipulate grain size in BCP systems. Controlled cooling through $\mathrm{T_{odt}}$ is one option, though variations of cooling rate become convoluted with alignment kinetics in field alignment studies. Adding a heterogeneous component to the system (\textit{i.e.} doping with a nucleating agent to decrease average grain size) can also be effective, particularly if the added species is colloidal, rather than molecular, in nature and therefore not likely to affect the interaction parameter $\chi$. The approach used here, blending, provides a third method, although we lack a clear interpretation of the reasons for its effectiveness. Blending produced samples with grain sizes that were intermediate between those of the neat BCPs without the addition of a chemically dissimilar component. The blended BCPs studied here display $\mathrm{T_{odt}}$ and d-spacings that interpolate between the neat blocks. The orientational order parameters vary concomitantly, suggesting that blending can be used as an effective strategy to modify grain growth kinetics in BCPs.

\begin{acknowledgements}

The authors thank N. Balsara and Z-G. Wang for insightful discussions regarding grain growth in block copolymers, S. Sinha for assistance with MALDI-TOF measurements, and the reviewers for helpful feedback. This work was supported by NSF under DMR-1410568 and DMR-1119826. Facilities use was supported by YINQE. P.G. acknowledges support under DMR-1507409. Additionally, this research used resources of the Center for Functional Nanomaterials, which is a U.S. DOE Office of Science Facility, at Brookhaven National Laboratory under Contract No. DE-SC0012704.
\end{acknowledgements}


\vspace{5mm}
{\bf\large{SUPPORTING INFORMATION}}

\section*{PS-b-P4VP Sample Information}

PS-b-P4VP 3.6K and 5.5K samples were synthesized by living anionic polymerization. The 3.6K material was synthesized in-house, while the 5.5K material was purchased from Polymer Source. The three blends of the two polymers were prepared in different ratios by dissolution of both in THF, dropcasting onto a heated slide at 35$^{\circ}$C, and subsequent vacuum annealing above the glass transitions of both blocks at 200 $^{\circ}$C for one hour to ensure complete solvent removal.

\begin{center}
\begin{table*}[ht]
 \begin{tabular}{||c| c| c| c|c|c||}
 \hline
Sample & Source & Mn(BCP) (kg/mol) & Mn(PS) (kg/mol)& Mn(P4VP) (kg/mol)& PDI \\ [0.5ex]
 \hline\hline
 3.6K & Gopalan Group & \textsuperscript{a}3.6 & \textsuperscript{a}1.9 &\textsuperscript{a}1.7 & \textsuperscript{c}1.04 \\
 \hline
 \multirow{2}{*}{5.5K} & \multirow{2}{*}{Polymer Source} & \textsuperscript{a}5.0 & \textsuperscript{a}2.4 & \textsuperscript{a}2.6 & \textsuperscript{c}1.04 \\
  & &\textsuperscript{b}5.5 & \textsuperscript{b}2.7 & \textsuperscript{b}2.8 & \textsuperscript{b}1.20   \\
 \hline
\end{tabular}
\end{table*}
\end{center}

\indent \textsuperscript{a}Mn determined via NMR by Gopalan Group at the University of Wisconsin-Madison

\textsuperscript{b}Mn and PDI information provided by Polymer Source

\textsuperscript{c}PDI calculated from MALDI-TOF-MS spectra

\section*{Determination of $\mathrm{T_{odt}}$}$\mathrm{T_{odt}}$ was determined from scattering data in the conventional manner as the temperature at which the system displayed a decrease in scattered intensity equivalent to 50\% of the intensity change on transiting the ODT during cooling.

\section*{Analysis of polydispersity by MALDI-TOF-MS spectra}

Gel permeation chromatography (GPC) could not be used effectively to compare the polydispersity indices (PDI) of the neat diblocks due to their small and similar values of their molecular weights. MALDI-TOF-MS was used instead. Samples for MALDI-TOF-MS analysis were prepared by mixing the polymer and matrix (2,5-dihydroxybenzoic acid)  in THF at a 10:1 ratio with an overall concentration of $\approx$ 0.03 mg/ml. 0.5 ul solutions were pipetted onto a MALDI plate, and spectra were collected in linear positive mode at an accelerating voltage of 25,000 V. $M_n$ and $M_w$ were calculated according to Equations (\ref{Mn}) and (\ref{Mw}). For each peak $i$, $M_i$ is the mass at the peak maximum and $A_i$ is the area for each mass.
\begin{eqnarray}
\label{Mn}
M_n=\frac{\Sigma A_i M_i}{\Sigma A_i} \\
\label{Mw}
M_w=\frac{\Sigma A_i {M_i}^2}{\Sigma A_i M_i}
\end{eqnarray}

\begin{figure}[!ht]

\includegraphics [width=80mm,scale=1]{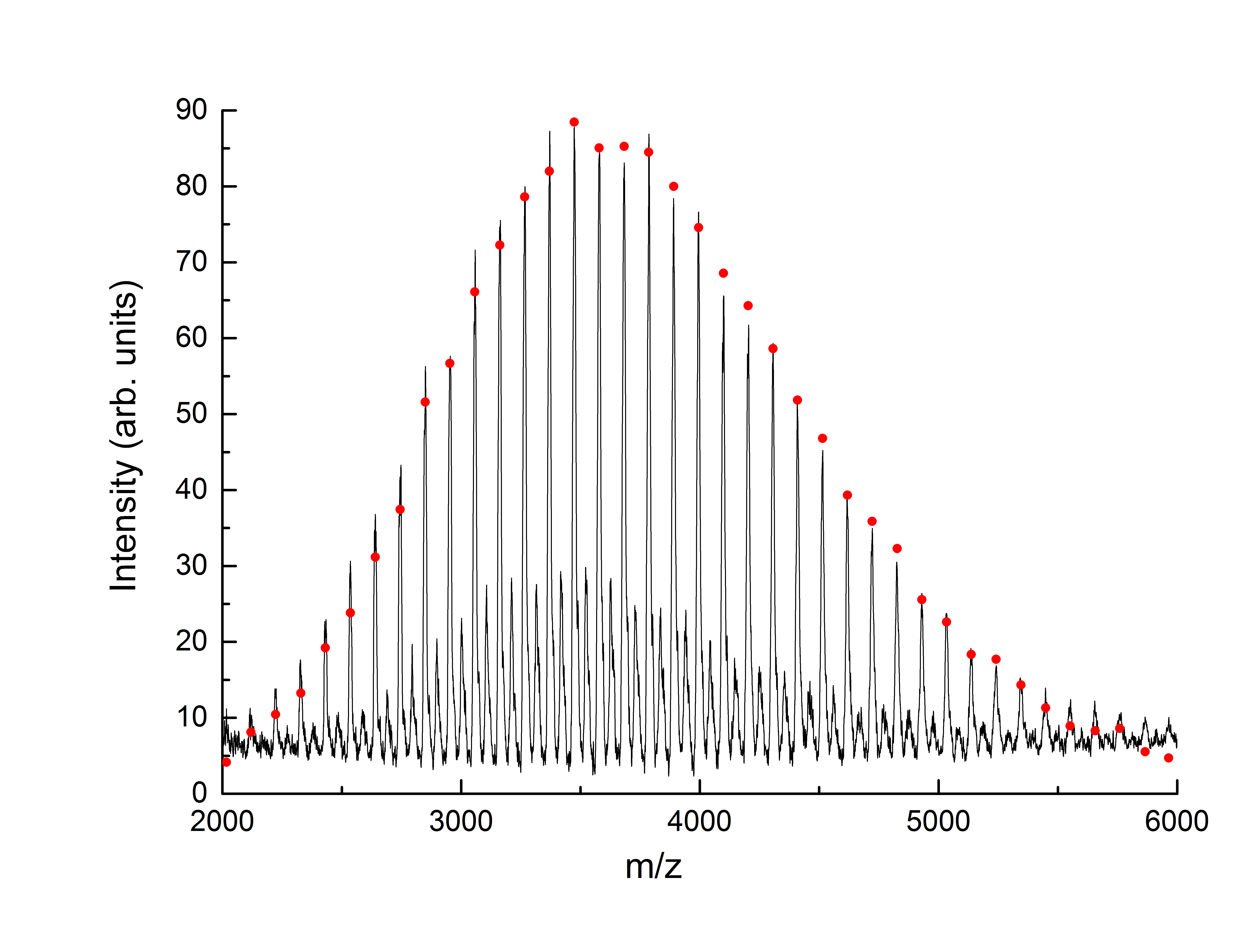}
  \caption{MALDI spectrum for 3.6K sample. $M_n=3771$ and $M_w=3932$. Red dots represent the relative probability for each curve, based on mass area.}
  \label{Maldi3_6K}
\end{figure}

The resulting MALDI spectra for samples 3.6K and 5.5K are shown below:

\begin{figure}[!ht]
\includegraphics [width=80mm, scale=1]{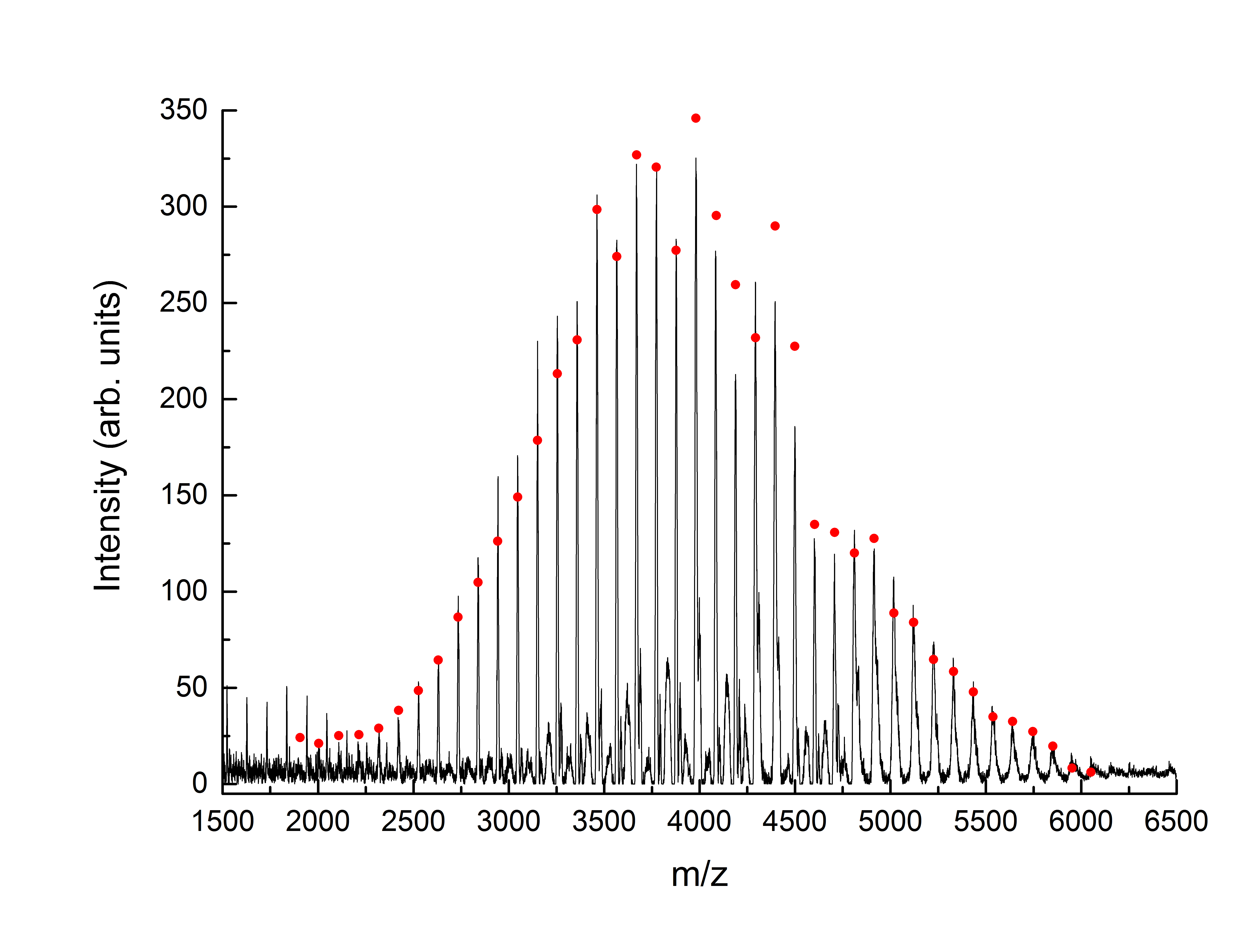}
  \caption{MALDI spectrum for 5.5K sample. $M_n=3903$ and $M_w=4048$. Red dots represent the relative mass probability for each curve, based on mass area.}
  \label{Maldi5_5k}
\end{figure}

\begin{figure}[!htb]
\includegraphics [width=80mm, scale=1]{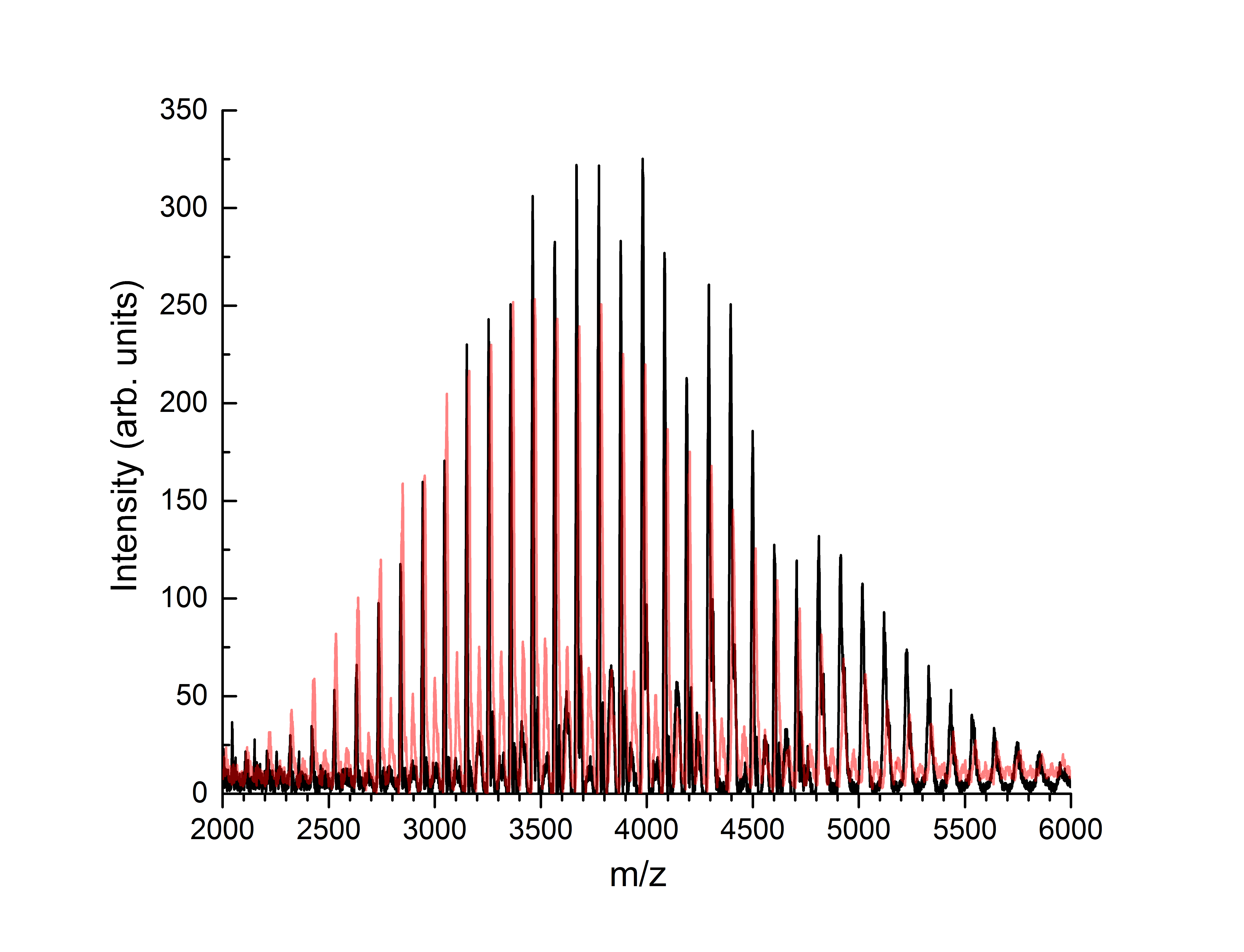}
  \caption{Overlay of 3.6K (red trace) and 5.5K (black trace) MALDI spectra.}
  \label{Overlayed}
\end{figure}

\clearpage

\section*{Representative TEM Images}

\begin{figure*}[!ht]
\includegraphics [width=110mm]{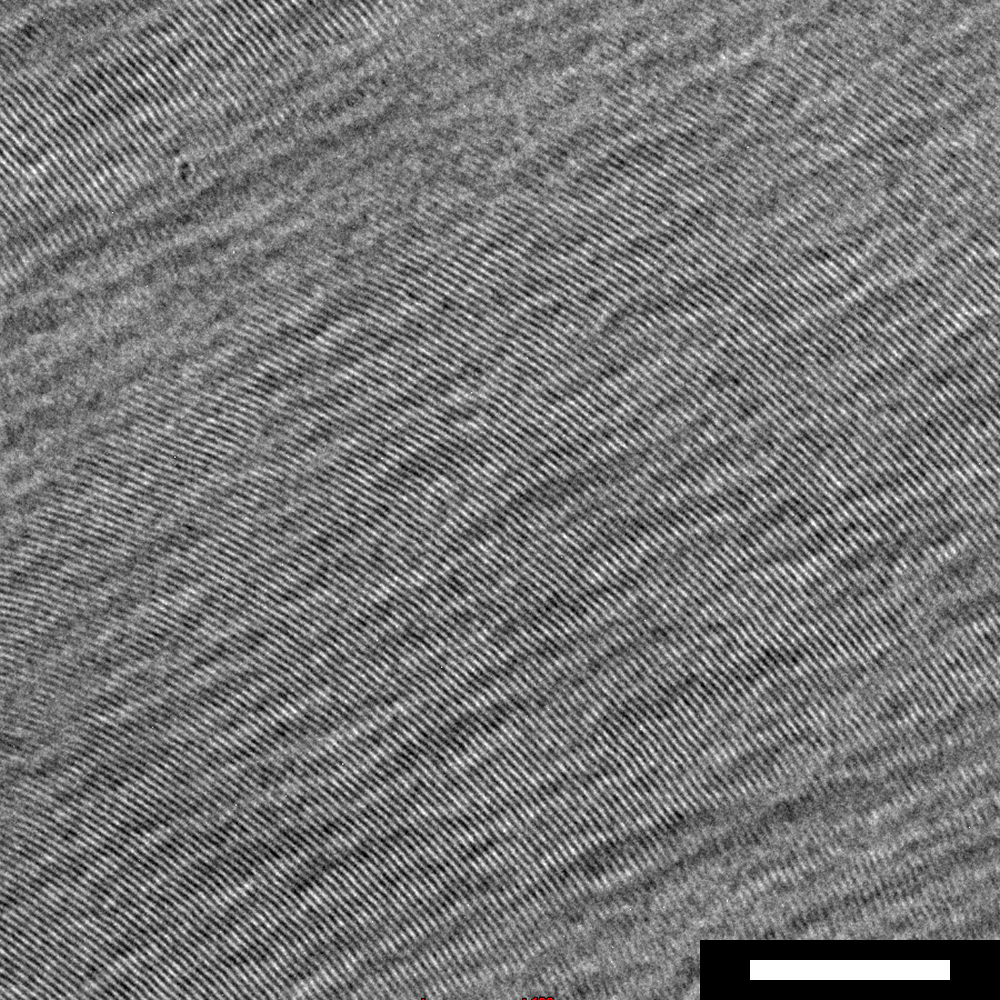}
  \caption{TEM image of PS-b-P4VP (3.6K) cooled through ${\textrm{T}_{\textrm{odt}}}$ in the absence of the field at 0.3 K/min. d-spacing=7.9 nm as determined by SAXS. Scale bar: 200nm.}
  \label{TEM3_6}
\end{figure*}

\clearpage

\begin{figure*}[!ht]
\includegraphics [width=110mm]{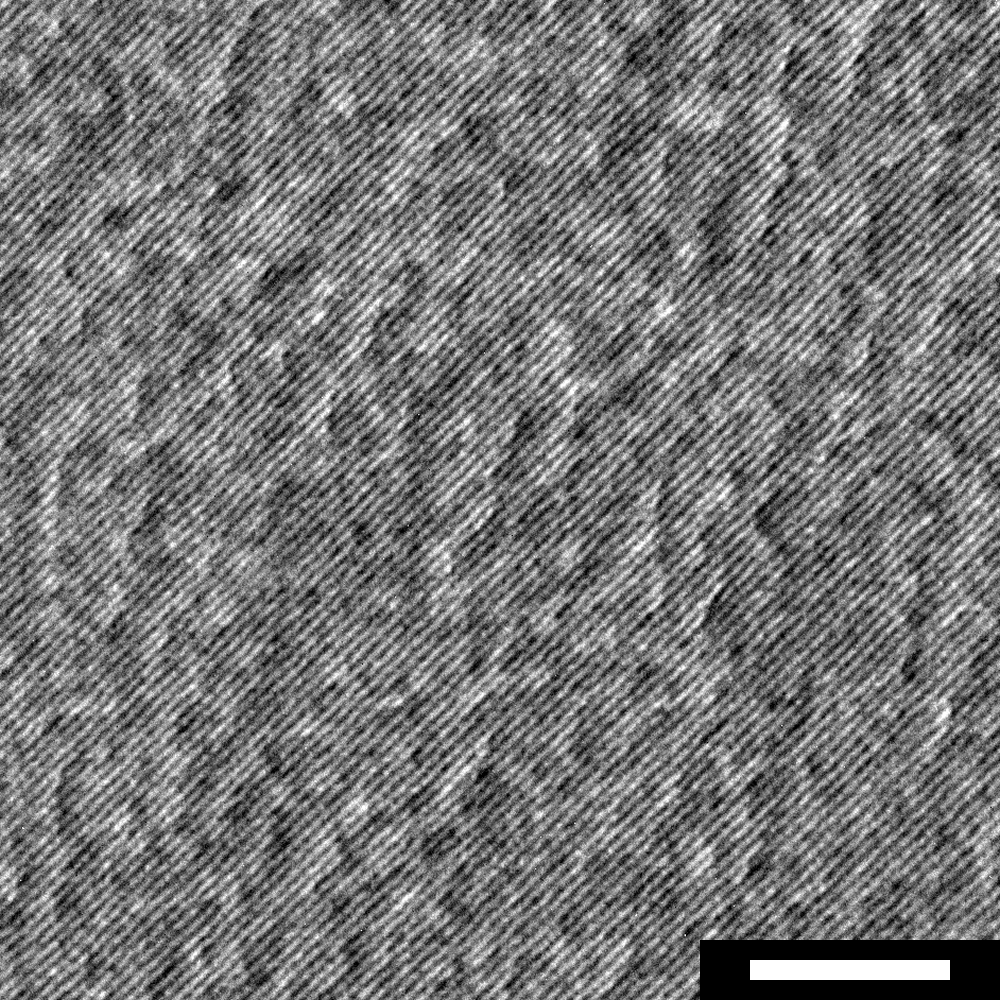}
  \caption{TEM image of PS-b-P4VP (5.5K) cooled through ${\textrm{T}_{\textrm{odt}}}$ in the absence of the field at 0.3 K/min. d-spacing=9.5 nm as determined by SAXS. Scale bar: 200nm.}
  \label{TEM5_5}
\end{figure*}
\clearpage
\newpage

\section*{Grain Size Determination by Variance Analysis}

\begin{figure}[h]
\includegraphics [width=80mm]{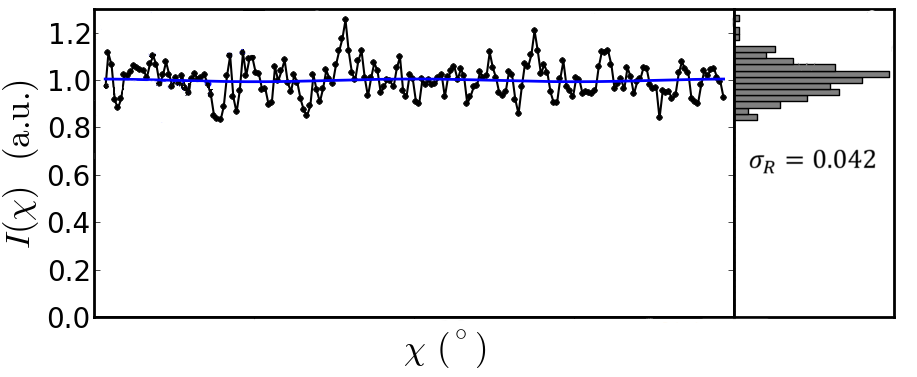}
  \caption{$I(\chi)$ vs. $\chi$ data from one measurement of the 3.6K sample. The left graph shows the integrated intensity along the scattering ring. The right graph shows a histogram of these intensity values, from which first $\sigma_R$, and then $N_g$ can be extracted.}
  \label{VarFigure}
\end{figure}

The samples were prepared as described in Ref.\cite{Rokhlenko2015}. Briefly, the samples used for this analysis were cooled 0.3 K/min. in the absence of the field and then mechanically polished to a thickness of 100-200 $\mu$m. The thickness for each was recorded to $\pm$ 2 $\mu$m accuracy. SAXS data was then collected in transmission mode for one hour from independent spots. A comprehensive description of the `variance scattering' method for grain size determination can be found in Ref.\cite{Yager2014}.
Briefly, each 2D scattering pattern is integrated along the scattering ring to give  $I(\chi)$ as shown in the left of Figure \ref{VarFigure}. Here, as an example, we show $I(\chi)$ for one measurement of the 3.6K sample. The right graph of Figure \ref{VarFigure} shows a histogram of these intensity values.

The standard deviation in these values $\sigma_R$ is one of four metrics that can be extracted from the raw data which can be used to calculate the number of grains probed $N_g$. The scaling of $\sigma_R$ and the other three metrics with respect to the number of grains $N_g$ is shown in Figure 3 of Ref.\cite{Yager2014}. All metrics gave consistent results, but for the results presented in the main text, we used the standard deviation metric $\sigma_R$. From Figure 3 of Ref.\cite{Yager2014}, we can see that $\sigma_R$ provides a clean power law scaling for $N_g$ over many orders of magnitude in the form $cN_g^{\beta}$:

\begin{equation}
\sigma_R=20.4 N_g^{-0.5}
\end{equation}

 With knowledge of the probed scattering volume (V) and extracted value $N_g$, the average characteristic grain dimension can be calculated  as $\xi=(V/N_g)^{1/3}$. This analysis implicitly assumes an isotropic distribution of grain orientations, with relatively well-defined grain boundaries. Twelve independent measurements on each sample were conducted to improve statistics, and to estimate measurement error bars.

\bibliographystyle{apsrev}
\bibliography{blends_arXiv}
\end{document}